\documentclass[prl,showpacs,amsmath,amssymb, twocolumn]{revtex4}
\usepackage{graphics}

\begin{document}


\title{Causal association of electromagnetic signals using the Cayley Menger determinant}%

\author{Samuel Picton Drake}
\affiliation{Defence Science and Technology Organisation, Edinburgh, South Australia 5111, Australia \\and \\ School of
Chemistry and Physics, The University of Adelaide, Adelaide, South Australia 5005, Australia}

\author{Brian D. O. Anderson}%
\affiliation{Research School of Information Sciences and Engineering, Australian National University, Canberra,
Australian Capital Territory 0200, Australia \\and\\ National ICT Australia (NICTA), Australian National
University, Canberra, Australian Capital Territory 0200, Australia}%
\author{Changbin Yu}
\affiliation{Research School of Information Sciences and Engineering, Australian National University, Canberra,
Australian Capital Territory 0200, Australia \\and\\ National ICT Australia (NICTA), Australian National
University, Canberra, Australian Capital Territory 0200, Australia}%


\begin{abstract}
In complex electromagnetic environments it can often be difficult to determine whether signals received by an antenna array emanated from the same source. The failure to appropriately assign signal reception events to the correct emission event makes accurate localization of the signal source impossible. In this paper we show that as the received signal events must lie on the light-cone of the emission event the Cayley–Menger determinate calculated from using the light-cone geodesic distances between received signals must be zero. This result enables us to construct an algorithm for sorting received signals into groups corresponding to the same far-field emission.
\end{abstract}

\maketitle

The deinterleaving of radar pulses is vital for the successful operation of radar warning receivers~\cite{wiley1982}. Modern radars can emit up to a million pulses per second, hence in a multi-emitter environment deinterleaving becomes a significant problem. Deinterleaving can be simplified if radars are distinguishable on the basis of other known pulse parameters such as transmitted power, carrier frequency or pulse width, however in many situations the radar pulse parameters are unknown or vary in an unpredictable way. There are a number of algorithms that are quite successful at deinterleaving pulse trains if the pulse repetition interval (PRI) is constant and a sufficiently long set of adjacent pulses is recorded~\cite{Conroy1998,Logothetis1998,Moore1994,milojevic1992}. Modern civilian marine radar however often have a pseudo-random PRI pattern to minimise the possible effects of interference; hence the techniques described in the above references will be ineffectual in deinterleaving the received radar pulses. In this paper we propose an alternative method for deinterleaving received radar pulses that requires no knowledge of the emitter radar characteristics and can be applied to only several pulses; this method is based on the concept of the light-cone used to describe events in Minkowski space-time~\cite{Misner1973}.

While we use radar pulses as an example in this letter it is important to realise that the proposed algorithm is valid for associating \emph{any} events that lie on the same light-cone. Furthermore although the following treatment assumes four receivers in $2+1$ space-time it can be extended trivially to five receivers in $3+1$ space-time.

The Cayley-Menger determinant for four points in Euclidean space is defined as~\cite{Blumenthal1953}
\begin{equation}\label{eq:detMplane4points}
D \equiv \det\left[
\begin{array}{ccccc}
0&s^2_{12}&s^2_{13}&s^2_{14}&1\\
s^2_{21}&0&s^2_{23}&s^2_{24}&1\\
s^2_{31}&s^2_{32}&0&s^2_{34}&1\\
s^2_{41}&s^2_{42}&s^2_{43}&0&1\\
1&1&1&1&0
\end{array}
\right]
\end{equation}
where $s^2_{ij}$ is the squared Euclidean distance between any two points in the space.

It has previously been noted that the Cayley-Menger determinant can be used to determine if a surface is \emph{flat}~\cite{weinberg1972}.
The Cayley-Menger determinate for a flat surface is zero if the distance between points on the surface is the geodesic distance which, in general, is not the Euclidean distance. The proof of this more general result can be verified by following the proof in Euclidean geometry as presented in section 3.1 of~\cite{michelucci2004} and realising that the result generalises so long as there exists a coordinate system in which the metric tensor has constant coefficients, which is an equivalent definition of the flatness of a surface~\cite{Misner1973}. It is immediately apparent from this definition that the Cayley-Menger determinant is zero in $2+1$ Minkowski space-time if the space-time events lie on a plane. If the Minkowski metric is used to calculate the space-time interval then $s_{ij}^2$ may be positive or negative depending on the signature of the Minkowski metric.

As the Cayley-Menger determinant specified by~\eqref{eq:detMplane4points} only contains terms that are cubic in $s_{ij}^2$, for errorless measurements
\begin{equation}\label{eq:CMDtest}
D =  \begin{cases}
0 &   \text{If all points lie on a flat surface} \\
\mathcal{O}\left(s_{ij}^6\right) & \text{If any points do not lie on a flat surface}
\end{cases}
\end{equation}

In $2+1$ space-time the light-cone is a two dimensional surface defined by all the possible paths of a photon emitted at a particular space-time point. Using the condition stated in~\eqref{eq:CMDtest} we conclude that the Cayley-Menger determinant is zero for all space-time events that lie on the same light-cone and hence are causally connected to the same emission event.

Evaluation of the Cayley-Menger determinant using~\eqref{eq:detMplane4points} requires knowledge of the geodesic distance between points on the light-cone. Calculation of the geodesic distance requires determining the metric tensor for the embedded conical surface and integrating the corresponding geodesic equation~\cite{Misner1973}. This process can be simplified considerably by choosing an appropriate set of coordinates $\{\xi,\psi,\zeta\}$ so that the metric tensor is diagonal and constant on the surface of the light-cone. If this can be done then the geodesic distance is the Euclidean distance in these coordinates.
\begin{figure}[p]
\caption{The light-cone and the a geodesic curve imbedded in Cartesian space-time\label{fig:lightConeConicCoords}}
\includegraphics{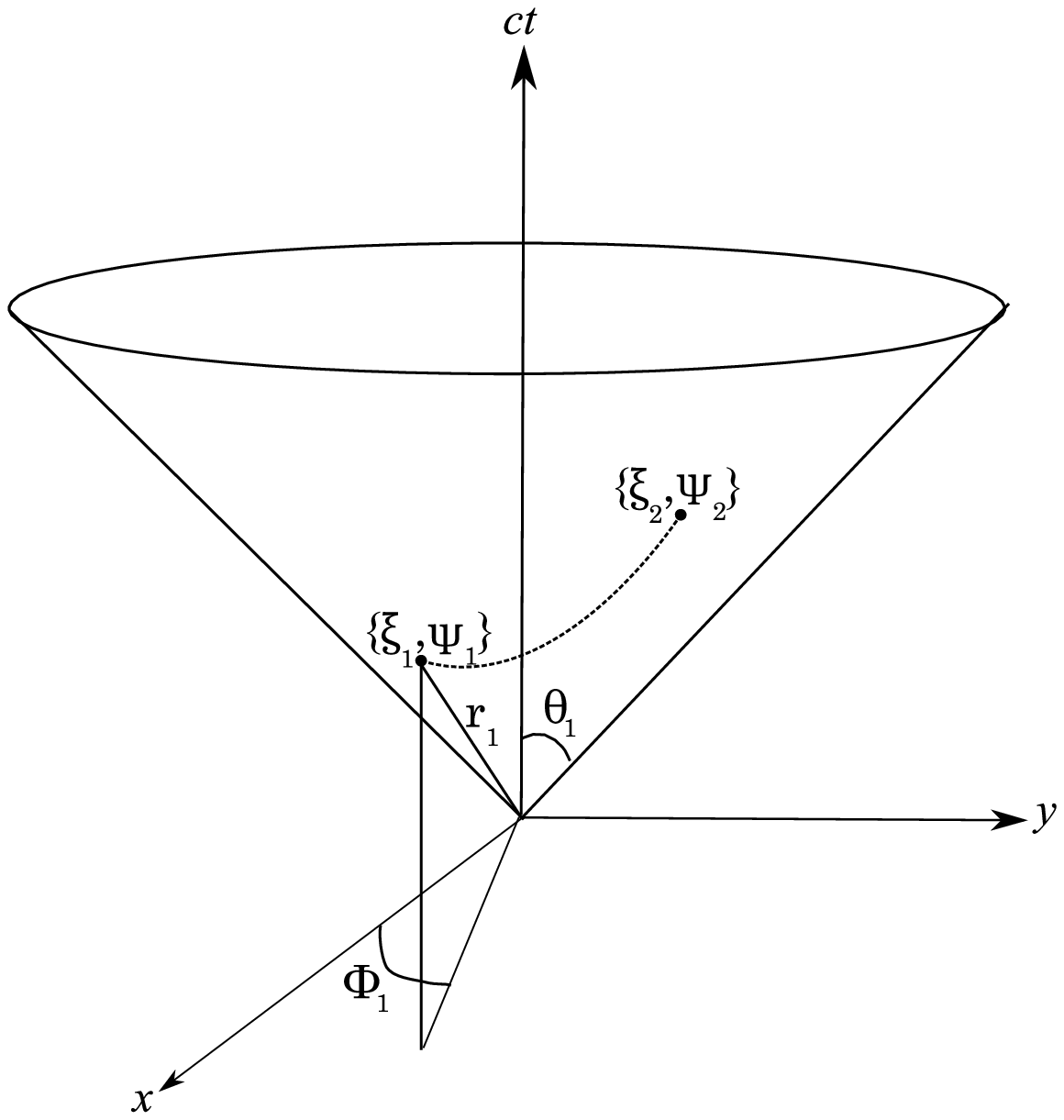}
\end{figure}

One such set of coordinates is given by
\begin{subequations}
\label{eq:conicalFnSpherical}
\begin{eqnarray}
\xi & = & r\cos\left(\phi\sin\theta\right) \\
\psi & = & r\sin\left(\phi\sin\theta\right) \\
\zeta  & = & \theta
\end{eqnarray}
\end{subequations}
and the corresponding inverse relations
\begin{subequations}
\label{eq:sphericalFnConical}
\begin{eqnarray}
r & = & \sqrt{\xi^2 + \psi^2}\\
\phi  & = & \frac{1}{\sin\zeta}\arctan\left(\frac{\psi}{\xi}\right) \\
\theta & = & \zeta
\end{eqnarray}
\end{subequations}
where $\{r,\theta,\phi\}$ are the spherical polar coordinates and are related to the $2+1$ cartesian space-time coordinates by
\begin{subequations}
\label{eq:SphericalFnCartesian}
\begin{eqnarray}
r & = & \sqrt{(x-x_0)^2 + (y-y_0)^2 + (t-t_0)^2} \\
\phi & = & \arctan\left(\frac{y-y_0}{x-x_0}\right) \\
\theta & = & \arctan\left(\frac{t-t_0}{\sqrt{(x-x_0)^2 + (y-y_0)^2}}\right)
\end{eqnarray}
\end{subequations}
where time is measured in natural units, i.e., $c=1$ and $\left[t_0, x_0, y_0\right]$ is the space-time coordinate of the emission event. Note that the coordinates defined by~\eqref{eq:conicalFnSpherical} are not the conical coordinates as defined by~\cite{moon1988}, they have been formulated so that the metric tensor on the cone is diagonal and constant.

The infinitesimal Euclidean distance between two points in coordinates defined by~\eqref{eq:conicalFnSpherical} is
\begin{eqnarray*}
ds^2 & = & d\xi^2 + d\psi^2
    +\left(\xi^2 + \psi^2\right)\left(1+\arctan\left(\frac{\psi}{\xi}\right)^2\cot\zeta^2\right) d\zeta^2 \nonumber \\
&&  + 2\psi\arctan\left(\frac{\psi}{\xi}\right)\cot\zeta d\xi d\zeta  \\
&&- 2\xi\arctan\left(\frac{\psi}{\xi}\right)\cot\zeta d\psi d\zeta
\end{eqnarray*}
If we constrain the path between any two points to be on the light-cone so that $\zeta = \frac{\pi}{4}$ then the geodesic distance between any two points $\{\xi_i,\psi_i,\zeta_i=\frac{\pi}{4}\}$ and $\{\xi_j,\psi_j,\zeta_j=\frac{\pi}{4}\}$ is
\begin{equation}\label{eq:OnLightConeDistance}
    s^2_{ij} = \left(\xi_i-\xi_j\right)^2+\left(\psi_i-\psi_j\right)^2
\end{equation}


The ultimate aim of many signal association algorithms is to provide information so that the emitter may be localised. Calculation of the Cayley-Menger determinant~\eqref{eq:detMplane4points} using~\eqref{eq:OnLightConeDistance} requires the space-time coordinates of the emission event for insertion into~\eqref{eq:SphericalFnCartesian} and hence would not appear to be useful.

In the far-field limit the geodesic distance on the light-cone can be approximated by the Euclidean distance i.e.,
\begin{equation*}\label{eq:smallEuclideanDist}
\left(\xi_i-\xi_j\right)^2+\left(\psi_i-\psi_j\right)^2 \approx
\left(x_i-x_j\right)^2+\left(y_i-y_j\right)^2 + \left(t_i-t_j\right)^2
\end{equation*}
The far field approximation is valid if the difference between the geodesic distance on the light-cone and the Euclidean distance is much less than the Euclidean distance. It can be shown that this condition is equivalent to the inter-antenna spatial distances being much less than the spatial distance to the emitter, i.e.,
\begin{equation*}\label{eqn:farFieldCondition}
\sqrt{(x_i-x_0)^2 + (y_i-y_0)^2} \gg \sqrt{(x_i-x_j)^2 + (y_i-y_j)^2}
\end{equation*}
as in a typical radar application.

In realistic scenarios the measured signal time of arrival is noisy which means that the expected value of $D$ from~\eqref{eq:detMplane4points} will not be zero, even if all the points lie on a flat surface. To calculate the expected value of $D$ in the presence of noise we would add a noise term $\epsilon^t_i$ to each of the receive events, calculate the square distance as a function of the noiseless distances and the noise, substitute these distances into~\eqref{eq:detMplane4points} and obtain an expression for the Cayley-Menger determinant in terms of the noiseless space-time distances and $\epsilon_t$.

This rather complex procedure can be avoided and the effect of noise on~\eqref{eq:detMplane4points} can be approximated by a rather simple expression once it is realised the Cayley-Menger determinant contains only terms of the type $s_{ij}^2s_{kl}^2s_{mn}^2$ and hence, to lowest order, the effect of noise can be approximated by $s_m^4\sigma_t^2$, where $s^2_m$ is the maximum square interval between events and $\sigma_t^2$ is the variance in the time of arrival noise. Using this result it is possible to construct a hypothesis test:
\begin{equation*}
\begin{array}{lcl}
H_0 & : & \textrm{All recieve events are caused by the same}\\
& &  \textrm{emission event}\\
\end{array}
\end{equation*}
The decision rule we use to test between this hypothesis is
\begin{equation}\label{eq:hypothesisTest}
H_0 = \begin{cases}
\textrm{true} &   \text{If} \quad D \leq 2(1.96\sigma_t)^2s_m^4 \\
\textrm{false} &  \text{If} \quad D >  2(1.96\sigma_t)^2s_m^4
\end{cases}
\end{equation}
The factors $2$ and $1.96$ come from that fact that the time difference variance is twice the time variance and that $95\%$ of time of arrival measurements lie between $\pm1.96\sigma_t$ of the noiseless time of arrival.

An algorithm for associating space-time measurements to a particular far-field emission event in $2+1$ space-time is as follows:
\begin{enumerate}
  \item Select four signal reception events from a list of possibilities\label{select}
  \item Determine the Euclidean space-time interval between all combinations of event quartets
  \item Calculate the Cayley-Menger determinant using~\eqref{eq:detMplane4points}. Use this to test $H_0$ using~\eqref{eq:hypothesisTest}\label{test}.
If $H_0$ is true then the events chosen in step~\ref{select}. are associated to the same emission event.
  \item Repeat steps~\ref{select} to \ref{test} until combinations are exhausted
\end{enumerate}

As an example of how this algorithm would work in practice consider a four element antenna array with four antennas placed at the four corners of a three metre square, i.e.,
\begin{equation}
{\bf p}_1 =  \begin{bmatrix}3\\0\end{bmatrix}
{\bf p}_2 =  \begin{bmatrix}0\\3\end{bmatrix}
{\bf p}_3 =  \begin{bmatrix}-3\\0\end{bmatrix}
{\bf p}_4 =  \begin{bmatrix}0\\-3\end{bmatrix}
\end{equation}
Consider a situation in which three signals arrive at each of the four antennas so that
\begin{subequations}\label{eq:simToA}
\begin{eqnarray}
t_1 & = & \{3.2400, 4.0133, 7.9263\} \\
t_2 & = & \{3.2334, 3.7789, 5.4521\} \\
t_3 & = & \{2.0474, 6.3819, 7.3851\} \\
t_4 & = & \{4.4990, 5.3704, 8.1967\}
\end{eqnarray}
\end{subequations}
How many emissions are there and what times are associated with each emission?

The times given in~\eqref{eq:simToA} were generated by adding zero mean Gaussian noise with 50~picosecond standard deviation to the arrival time of signals generated by two emitters located 5km from the centre of the antenna array. The third set of arrival times was generated by a random number generator and could represent spurious measurements. There are $3^4$ ways in which the sample data can be grouped into quartets, however of these 81 possible combinations only two satisfy the hypothesis test as stated in~\eqref{eq:hypothesisTest}. Using the algorithm outlined above we are able to correctly conclude that there are two emissions events whose corresponding receive events $\boldsymbol{x}=\left[ct,x,y\right]$ are
\begin{equation*}
\begin{array}{lcl}
\boldsymbol{x}_1 & = &   \left[4.0133,3,0\right]^T\\
\boldsymbol{x}_2 & = &   \left[3.2334,0,3\right]^T\\
\boldsymbol{x}_3 & = &   \left[7.3851,-3,0\right]^T\\
\boldsymbol{x}_4 & = &   \left[8.1967,0,-3\right]^T
\end{array}
\textrm{and}
\begin{array}{lcl}
\boldsymbol{x}_1 & = &   \left[7.9263,3,0\right]^T\\
\boldsymbol{x}_2 & = &   \left[5.4521,0,3\right]^T\\
\boldsymbol{x}_3 & = &   \left[2.0474,-3,0\right]^T\\
\boldsymbol{x}_4 & = &   \left[4.4990,0,-3\right]^T
\end{array}
\end{equation*}
Each quartet of events can used to estimate the location emitter using the time difference of arrival technique~\cite{Torrieri1984}.

The Defence Science and Technology Organisation (DSTO) of Australia is building an antenna array system to test the association algorithm outlined in this paper. The antenna array will be constructed using a precise time interval measuring unit such as the ATMD-GPX~\cite{acam2005}. Once the evaluation system is built trials will be conducted with two and three radars placed at about 5km from the antenna array. The tests will be conducted at the DSTO signals testing facility in Adelaide, Australia. Results of the trial will be reported elsewhere.
\newpage

\begin{thebibliography}{12}
\expandafter\ifx\csname natexlab\endcsname\relax\def\natexlab#1{#1}\fi
\expandafter\ifx\csname bibnamefont\endcsname\relax
  \def\bibnamefont#1{#1}\fi
\expandafter\ifx\csname bibfnamefont\endcsname\relax
  \def\bibfnamefont#1{#1}\fi
\expandafter\ifx\csname citenamefont\endcsname\relax
  \def\citenamefont#1{#1}\fi
\expandafter\ifx\csname url\endcsname\relax
  \def\url#1{\texttt{#1}}\fi
\expandafter\ifx\csname urlprefix\endcsname\relax\def\urlprefix{URL }\fi
\providecommand{\bibinfo}[2]{#2}
\providecommand{\eprint}[2][]{\url{#2}}

\bibitem[{\citenamefont{Wiley}(1982)}]{wiley1982}
\bibinfo{author}{\bibfnamefont{R.~G.} \bibnamefont{Wiley}},
  \emph{\bibinfo{title}{Electronic intelligence : the analysis of radar
  signals}} (\bibinfo{publisher}{Artech House}, \bibinfo{address}{Dedham, MA},
  \bibinfo{year}{1982}).

\bibitem[{\citenamefont{Conroy and Moore}(1998)}]{Conroy1998}
\bibinfo{author}{\bibfnamefont{T.}~\bibnamefont{Conroy}} \bibnamefont{and}
  \bibinfo{author}{\bibfnamefont{J.~B.} \bibnamefont{Moore}},
  \bibinfo{journal}{Signal Processing, IEEE Transactions on}
  \textbf{\bibinfo{volume}{46}}, \bibinfo{pages}{3326} (\bibinfo{year}{1998}).

\bibitem[{\citenamefont{Logothetis and Krishnamurthy}(1998)}]{Logothetis1998}
\bibinfo{author}{\bibfnamefont{A.}~\bibnamefont{Logothetis}} \bibnamefont{and}
  \bibinfo{author}{\bibfnamefont{V.}~\bibnamefont{Krishnamurthy}},
  \bibinfo{journal}{Signal Processing, IEEE Transactions on}
  \textbf{\bibinfo{volume}{46}}, \bibinfo{pages}{1344} (\bibinfo{year}{1998}).

\bibitem[{\citenamefont{Moore and Krishnamurthy}(1994)}]{Moore1994}
\bibinfo{author}{\bibfnamefont{J.~B.} \bibnamefont{Moore}} \bibnamefont{and}
  \bibinfo{author}{\bibfnamefont{V.}~\bibnamefont{Krishnamurthy}},
  \bibinfo{journal}{Signal Processing, IEEE Transactions on}
  \textbf{\bibinfo{volume}{42}}, \bibinfo{pages}{3092} (\bibinfo{year}{1994}).

\bibitem[{\citenamefont{Milojevic and Popovic}(1992)}]{milojevic1992}
\bibinfo{author}{\bibfnamefont{D.~J.} \bibnamefont{Milojevic}}
  \bibnamefont{and} \bibinfo{author}{\bibfnamefont{B.~M.}
  \bibnamefont{Popovic}}, \bibinfo{journal}{Radar and Signal Processing, IEE
  Proceedings F} \textbf{\bibinfo{volume}{139}}, \bibinfo{pages}{98}
  (\bibinfo{year}{1992}).

\bibitem[{\citenamefont{Misner et~al.}(1973)\citenamefont{Misner, Thorne, and
  Wheeler}}]{Misner1973}
\bibinfo{author}{\bibfnamefont{C.~W.} \bibnamefont{Misner}},
  \bibinfo{author}{\bibfnamefont{K.~S.} \bibnamefont{Thorne}},
  \bibnamefont{and} \bibinfo{author}{\bibfnamefont{J.~A.}
  \bibnamefont{Wheeler}}, \emph{\bibinfo{title}{Gravitation}}
  (\bibinfo{publisher}{W. H. Freeman and Company}, \bibinfo{year}{1973}).

\bibitem[{\citenamefont{Blumenthal}(1953)}]{Blumenthal1953}
\bibinfo{author}{\bibfnamefont{L.~M.} \bibnamefont{Blumenthal}},
  \emph{\bibinfo{title}{Applications of Distance Geometry}}
  (\bibinfo{publisher}{Oxford University Press}, \bibinfo{address}{Oxford},
  \bibinfo{year}{1953}).

\bibitem[{\citenamefont{Weinberg}(1972)}]{weinberg1972}
\bibinfo{author}{\bibfnamefont{S.}~\bibnamefont{Weinberg}},
  \emph{\bibinfo{title}{Gravitation and cosmology: principles and applications
  of the general theory of relativity}} (\bibinfo{publisher}{Wiley},
  \bibinfo{address}{New York}, \bibinfo{year}{1972}).

\bibitem[{\citenamefont{Michelucci and Foufou}(2004)}]{michelucci2004}
\bibinfo{author}{\bibfnamefont{D.}~\bibnamefont{Michelucci}} \bibnamefont{and}
  \bibinfo{author}{\bibfnamefont{S.}~\bibnamefont{Foufou}}, in
  \emph{\bibinfo{booktitle}{ACM Symposium on Solid Modeling and Applications}}
  (\bibinfo{publisher}{ACM Press}, \bibinfo{address}{Genova, Italy},
  \bibinfo{year}{2004}), pp. \bibinfo{pages}{285--290}.

\bibitem[{\citenamefont{Moon and Spencer}(1988)}]{moon1988}
\bibinfo{author}{\bibfnamefont{P.}~\bibnamefont{Moon}} \bibnamefont{and}
  \bibinfo{author}{\bibfnamefont{D.~E.} \bibnamefont{Spencer}},
  \emph{\bibinfo{title}{Field theory handbook : Incl. coordinate systems,
  differential equations and their solutions}} (\bibinfo{publisher}{Springer},
  \bibinfo{address}{Berlin}, \bibinfo{year}{1988}), \bibinfo{edition}{3rd} ed.

\bibitem[{\citenamefont{Torrieri}(1984)}]{Torrieri1984}
\bibinfo{author}{\bibfnamefont{D.~J.} \bibnamefont{Torrieri}},
  \bibinfo{journal}{Aerospace and Electronic Systems, IEEE Transactions on}
  \textbf{\bibinfo{volume}{AES-20}}, \bibinfo{pages}{183}
  (\bibinfo{year}{1984}).

\bibitem[{\citenamefont{acam~mess electonic}(2005)}]{acam2005}
\bibinfo{author}{\bibnamefont{acam~mess electonic}},
  \emph{\bibinfo{title}{Atmd-gpx tdc-gpx evaluation system: Datasheet}},
  \bibinfo{howpublished}{World Wide Web electronic publication}
  (\bibinfo{year}{2005}),
  \urlprefix\url{http://www.acam.de/fileadmin/Download/pdf/English/DB_AMGPX_e.%
pdf}.

\end{thebibliography}

\end{document}